\documentclass[aps,prd,showpacs,preprintnumbers,nofootinbib,floatfix,floats,groupedaddress,twocolumn]{revtex4}
\usepackage{bm}
\usepackage{latexsym}
\usepackage{dcolumn}
\usepackage{amsmath,amsfonts,amssymb}
\usepackage{graphicx,epsfig}
\usepackage{color}
\def\nn{\nonumber}
\def\eq#1{{Eq.~(\ref{#1})}}
\def\l{\left}
\def\r{\right}
\def\DM{\mathrm{d}}

\reversemarginpar

\begin{document}
\title{Response of Unruh-DeWitt detector with time-dependent acceleration}

 \author{Dawood Kothawala}
 \email{dawood@iucaa.ernet.in}
 \author{T.~Padmanabhan}
 \email{paddy@iucaa.ernet.in}
 \affiliation{IUCAA,
 Post Bag 4, Ganeshkhind, Pune - 411 007, India}

\date{\today}
\begin{abstract}
It is well known that a detector, coupled linearly to a quantum field and accelerating
through the inertial vacuum with a constant acceleration $g$, will behave as though it is
immersed in a radiation field with temperature $T=(g/2\pi)$. We study a generalization of this result for detectors moving with a time-dependent acceleration $g(\tau)$ along
a given direction. After defining the rate of excitation of the detector appropriately,
we evaluate this rate for time-dependent acceleration, $g(\tau)$, to linear order in the parameter $\eta = \dot g / g^2$.
In this case, we have three length scales in the problem:
$g^{-1},\ (\dot g/g)^{-1}$ and $\omega^{-1}$
where $\omega$ is the energy difference between the two levels of the detector at which the spectrum is probed. We show that:
(a) When $\omega^{-1} \ll g^{-1} \ll (\dot g/g)^{-1}$, the rate of transition of the
detector  corresponds to a slowly varying temperature $T(\tau) = g(\tau)/2 \pi $, as one would have expected.
(b) However, when $ g^{-1}\ll \omega^{-1}  \ll (\dot g/g)^{-1}$, we find that
 the spectrum is modified \textit{even at the order $O(\eta)$}.
 This is counter-intuitive because, in this case,
 the relevant frequency does not probe the rate of change of the acceleration since
 $(\dot g/g) \ll \omega$ and we certainly do not have deviation from the thermal
 spectrum when $\dot g =0$. This result shows that there is a subtle discontinuity
 in the behaviour of detectors with $\dot g = 0$ and $\dot g/g^2$ being arbitrarily small.
 We  corroborate this result by evaluating the detector response for a particular trajectory which admits an analytic expression for the poles of the Wightman function.
\end{abstract}

\maketitle
\vskip 0.5 in
\noindent
\maketitle
\section{Introduction} \label{sec:intro}
One of the key results which emerge from the study of quantum field theory in non-inertial coordinate systems (and curved spacetime though we will not consider it in this paper)
is that both the particle content of the quantum states, as well as the pattern of
vacuum fluctuations, are not generally covariant. This can be explicitly demonstrated by studying the response of detectors linearly coupled to the quantum field (usually called Unruh-DeWitt detectors) in different states of motion (\cite{unruh,birrell-1982}; see \cite{crispino-review} for a review).
The probability that a detector traveling along the trajectory $X_*^i (\tau)$
will get excited can be expressed as an integral over the Wightman function of the
field in the form
\begin{eqnarray}
\mathcal{P} &=& \int_{-\infty}^{\infty} \DM \tau_2 \int_{-\infty}^{\infty} \DM \tau_1 \exp{\l( -i \omega u \r)} \; G^{+}[X^i_{*}(\tau_2), X^j_{*}(\tau_1)]
\nn \\
\nn \\
&=& \int_{-\infty}^{\infty} \DM t \int_{-\infty}^{\infty} \DM u \exp{\l( -i \omega u \r)} \; G^{+}[u,t]
\label{response1}
\end{eqnarray}
where $\omega$
is the energy difference between the two levels and
the coordinates $(u, t)$ are introduced through the definitions: $u = \tau_2 - \tau_1$, $2 t = \tau_2 + \tau_1$, and we have absorbed a factor dependent upon the internal details of the detector in the definition of $\mathcal{P}$.

When the trajectory $X_*^i (\tau)$ is along the integral curve of a time-like
Killing vector field in flat spacetime (we will call such trajectories `stationary'), the Wightman function $G^+[u,t]$
will only depend on the time difference $u$ so that $G^+[u,t]= G[u]$. Then
the second integral over $t$ in \eq{response1} will lead to a divergent result.
This is handled by the usual procedure of time dependent perturbation theory
which involves ignoring the integral over $t$ and interpreting
the rest of the result as providing the rate of transitions between the two levels.
For the stationary trajectories, this rate will be a constant.

It is also worth mentioning at this point that the rate so defined is a real number. This is easily seen by
noting that, the imaginary part of the rate will be given by:
\begin{eqnarray}
\mathrm{Im}[\mathcal{\dot P}] = \frac{1}{2 i} \int_{-\infty}^{\infty} \DM u &\Biggl\{& \exp{\l( -i \omega u \r)} \;
G^{+}[u,t]
\\ \nonumber
&-& \exp{\l( i \omega u \r)} \; (G^{+})^*[u,t] \Biggl\}
\end{eqnarray}
We now note that the Hadamard function satisfies: $(G^{+})^*[P,Q] = G^+[Q,P]$ for any two points $(P, Q)$,
here characterised by $(\tau_1, \tau_2)$. Further, it follows from the definitions of $u$ and $t$ that interchanging
the points amounts to $u \rightarrow -u$ and $t \rightarrow t$, so that $(G^{+})^*[u,t] = G^{+}[-u,t]$. It is then easy
to see that Im$[\mathcal{\dot P}] = 0$.

To avoid possible confusion, we must also point out that our definition for the
response function differs from certain other definitions found in literature (often motivated by arguments
of causality etc.). Essentially, the difference lies in the choice of the ``time" variable with respect to which the rate is
defined (which, in our case, is $t$). Similarly, our choice of regularization scheme is also different from some
other choices found in literature. Hence, the result we shall obtain can not be directly compared with other
results based on a different choice of definition for the rate or regularization scheme. We shall have more to say on
this in the last section.

Since flat spacetime admits ten independent Killing vector fields, one can construct
several linear combinations of these Killing vectors which will be time-like in parts of
the spacetime. The response of detectors on these trajectories have been studied extensively in the literature (see, e.g., Ref. \cite{tpapsc,letaw,probe}).
In a generic situation, the detector will respond to the pattern of vacuum fluctuations
which can coincide with the particle content of the quantum field determined by
Bogoluibov coefficients in specific cases but not always.
( e.g., for a circular trajectory, the particle detector ``clicks" but the number of particles calculated using the Bogoluibov coefficients turns out to be zero.
Many of these conceptual issues have been discussed and clarified in the
literature \cite{probe,tpclick,tptps}.)

The uniformly accelerated trajectory corresponds to the integral curve of the Killing vector field corresponding to the Lorentz boost along the direction of the acceleration $g$.
In this case \cite{comment} we obviously have $G^+[u,t]= G[u]$.
In this particular case, the pattern of vacuum fluctuations match with
the particle content of the quantum state and the rate of excitation of the detector
will correspond to a thermal spectrum of particles with a temperature $T=g/2\pi$.
This is of particular importance because it allows us to associate a temperature with the Rindler horizon with obvious implications for black hole physics.

Unfortunately, a detector which is uniformly accelerated from $\tau = -\infty$ to
$\tau = + \infty$ is not physically realizable. The question arises as to what happens in the case of more realistic detectors. One possible way of addressing this question is to keep the coupling to the field switched on only for a finite interval of time (see, e.g,  Ref. \cite{sriram-finite-time}).
But this introduces transients and one needs to handle them with care. It also does not seem very natural to switch off the coupling in this manner. A more obvious and physically interesting way of attacking the problem would be to study the response of a detector
moving along a given direction with a  time-dependent acceleration $g(\tau)$ which is
what we will do in this paper.

There are three further
motivations for taking up this study which are somewhat indirect.

First, we know that
there is a direct correspondence between the detector response in a uniformly accelerated trajectory and the phenomena which takes place in the Hartle-Hawking vacuum state
around a black hole. By extending this analogy, we would expect a sub-class of
time-dependent accelerations --- especially those $g(\tau)$ which vanish at early
times and become constant at late times --- to correspond to the phenomena which takes place
in a collapsing black hole scenario in the Unruh vacuum state.
(For preliminary discussions along these lines, see section 5.1 of Ref. \cite{tppr}.)
This would be interesting to study.

Second, there has been considerable amount of work in recent years which attempts to
interpret the field equations of gravity as a thermodynamic identity. This body of work
\cite{tptedetc} uses the concept of local Rindler observers that corresponds to trajectories
which, in the local inertial frames around any given event, will be a hyperbola.
While one expects such a local concept to be valid as a first approximation,
it is important to verify it explicitly (and indeed our results in this paper will justify this notion and make it sharper).

Finally, this subject has thrown up fair number of surprises and subtleties in the
past and one cannot take it for granted that
intuitively obvious results will arise when we rigorously analyse the case of, say,
a slowly varying acceleration!
It requires  explicit verification.
Our naive expectation will be that, for sufficiently slowly varying acceleration (with
$(\dot g/g^2)\ll 1$) one would expect the detection rate to correspond to  a time dependent temperature $T(\tau)\propto g(\tau)$. At the same time, one will \textit{not} expect such a result to hold for all frequencies of the thermal spectrum.
There is, in fact, a good reason to expect some modification due the presence of (local acceleration) horizon. This sets a length scale $g^{-1}$ in the problem, which can be compared with the length scale probed by a particular mode, $\omega^{-1}$. Of course, we know that the spectrum is Planckian for all values of $g^{-1} \omega$ when $g$ is constant; it is therefore interesting to see whether a varying $g$ makes any difference.  As we shall show, one does get low frequency ($g^{-1} \omega \ll 1$) modifications when $\dot g$ is non-vanishing even when $\dot g/g\ll\omega$, which is a surprising result.

In Sec. \ref{sec:setup}, we describe the setup appropriate for calculating the response function. In Sec. \ref{sec:time-dep-acc}, we evaluate the Unruh-DeWitt detector response for time-dependent acceleration, $g(\tau)$, to linear order in the parameter $\eta = \dot g / g^2$. We find that, to this order, the spectrum can indeed be approximated in the UV region ($\omega \gg g $) by Planck spectrum with time-dependent temperature, $T= g(\tau)/2 \pi $. However, the spectrum is modified \textit{even at $O(\eta)$} for $\omega \ll g$. In Sec. \ref{sec:specific-acc}, we corroborate this result by evaluating the detector response for a particular trajectory which admits an analytic expression for the poles (under a particular approximation). Finally, we conclude with few relevant comments.
We use the metric signature $(-, +, +, +)$.
\section{Detector response: Background} \label{sec:setup}
The trajectory of an observer moving with a time-dependent acceleration, $g(\tau)$, with $\tau$ being the proper time, is given by
\begin{eqnarray}
T_{*}(\tau) &=& \int^{\tau} \DM \alpha \; \cosh{\chi(\alpha)} \nn \\
X_{*}(\tau) &=& \int^{\tau} \DM \alpha \; \sinh{\chi(\alpha)}
\label{eq:trajectory}
\end{eqnarray}
where
\begin{eqnarray}
\frac{\DM \chi(\tau)}{\DM \tau} = g(\tau)\, ; \qquad
\chi(\tau) = \int^{\tau}_0 \DM \alpha \; g(\alpha)
\end{eqnarray}
and $(X^0, X^1)=(T, X)$ are inertial coordinates. (We have chosen $\chi(0)=0$
to obtain the integral form.)  The local coordinates of the observer, $(\tau, x)$, can be constructed easily; these are given by (see, for e.g, Equation (73) of Ref. \cite{tppr})
\begin{eqnarray}
T(\tau) &=& \int^{\tau} \DM \alpha \; \l[ 1 + g(\alpha) x \r] \; \cosh{\chi(\alpha)} \nn \\
X(\tau) &=& \int^{\tau} \DM \alpha \; \l[ 1 + g(\alpha) x \r] \; \sinh{\chi(\alpha)}
\label{eq:local-coords}
\end{eqnarray}
In the local coordinates, the observer is always located at $x=0$. We shall assume $\tau_2 > \tau_1$ without  loss of generality.

The probability of transition for the detector is  given by \eq{response1}
which is valid for any trajectory.
In general, for an arbitrary $g(\tau)$, there is no time translational symmetry and $G^+$ will depend on both $u$ and $t$. Following the procedure adapted for stationary trajectories,  we shall again define the transition rate by
ignoring the integral over $t$. But now
this rate  will be time-dependent, due to the $t-$dependence of $G^+$, which, of course,
is to be expected. So, we shall define the transition rate to be
\begin{eqnarray}
\mathcal{\dot P} &=& \int_{-\infty}^{\infty} \DM u \exp{\l( -i \omega u \r)} \; G^{+}(u, t)
\label{eq:trans-rateF}
\end{eqnarray}
The Wightman function is given by
\begin{eqnarray}
G^{+}(1,2) &=& \frac{1}{4 \pi^2} \frac{1}{\ell^2}
\nn \\
\ell^2(1,2) &=& - \l[ T_{*}(\tau_2) - T_{*}(\tau_1) \r]^2 + \l[ X_{*}(\tau_2) - X_{*}(\tau_1) \r]^2
\nn \\
\label{eq:wightman-fun}
\end{eqnarray}
with an $i\epsilon$ prescription which is implicit in the difference of the time coordinates.
Substituting Eqs.~(\ref{eq:trajectory}), the expression for $\ell^2$ can be written in the following convenient form;
\begin{eqnarray}
\ell^2 = - I_{+} \; I_{-}
\label{eq:geodesic-length}
\end{eqnarray}
where
\begin{eqnarray}
I_{\pm} = \int_{\tau_1}^{\tau_2} \DM \alpha \; \exp{\pm \chi(\alpha)}
\label{eq:ipm-defs}
\end{eqnarray}
We see that the detector response is essentially determined by the poles of $\ell^2$. For the constant acceleration case, the poles of $I_+$ and $I_-$ coincide, so that we have an infinity of second order poles, the residues at which
gives the well known thermal response function.
Our task, therefore, reduces to identifying the poles of $I_\pm$ and evaluating the
integral in \eq{eq:ipm-defs}. We shall now turn to this task.

\section{Detector response for $\dot g / g^2 \ll 1$} \label{sec:time-dep-acc}

As one can easily see, it is impossible to determine the structure of the poles for a
general $g(\alpha)$. Hence, we shall attack this problem in two steps. First, in this
section, we will consider a slowly varying acceleration and obtain the detector response.
In the next section, we shall work out the response for a specific form of $g(\tau)$.

Consider a general $g(\tau)$, which varies slowly compared to its value $g_0$ at $\tau=0$ which can be chosen to be an arbitrary instant of proper time. We shall now expand $g(\tau)$ in a  Taylor series retaining the lowest order terms:
\begin{eqnarray}
g(\tau) &=& g_0 + \dot g_0 \tau + O(\ddot g_0 \tau^2) \nn \\
\chi(\tau) &=& g_0 \tau + \frac{1}{2} \dot g_0 \tau^2 + O(\ddot g_0 \tau^3) \nn \\
&\approx& g_0 \tau \l[ 1 + \frac{1}{2} \l( \frac{\dot g_0}{g_0} \r) \tau \r]
\end{eqnarray}
Therefore, we have
\begin{eqnarray}
\exp{\pm \chi(\tau)} = \exp{\left(\pm g_0 \tau \right)} \l[ 1 \pm \frac{1}{2} \eta_0 (g_0 \tau)^2 + O(\eta_0^2) \r]
\end{eqnarray}
where we have defined $\eta_0 = \dot g_0 / g_0^2$, and we shall do subsequent calculations keeping terms up to $O(\eta_0)$. A trajectory is, of course, not completely specified by $\eta_0$. For our result to remain valid, the contribution of higher derivatives of acceleration must be ignorable compared to $\dot g$. Although a restriction, this condition will almost always be fulfilled in physically relevant cases, when there is only one small parameter in the problem. If not we will get the same result when all the corresponding higher derivatives of the acceleration are small.

We now proceed to analyze the pole structure of $\ell^2$ to determine the detector response. Evaluation of $I_{\pm}$ involves trivial integrations; we obtain,
\begin{eqnarray}
I_{\pm} = \frac{1}{g_0} \l[ \l( 1 \pm \frac{1}{2} \eta_0 \frac{\DM^2}{\DM \alpha^2} \r) Q_{\pm}\l( \xi_1, \xi_2; \alpha \r) \r]_{\alpha=1}
\end{eqnarray}
where we have defined
\begin{eqnarray}
Q_{\pm}\l( \xi_1, \xi_2; \alpha \r) = \int^{\xi_2}_{\xi_1} \DM \xi \; \exp{\pm \alpha \xi}
\label{eq:Q-def}
\end{eqnarray}
with $\xi_{1 (2)} = g_0 \tau_{1 (2)}$. Therefore, we obtain,
\begin{eqnarray}
\frac{1}{I_+ I_-} \approx \frac{ g_0^2}{Q_+ Q_-} \l[ 1 + \frac{\eta_0}{2} \frac{1}{Q_+ Q_-} \l( Q_+ Q''_{-} - Q_- Q''_{+} \r) \r] \nn
\\
\label{eq:poles-1}
\end{eqnarray}
where $\approx$ sign implies that we have  ignored $O(\eta_0^2)$ terms, as we should for consistency. (The prime stands for $\DM / \DM \alpha$, with $\alpha$ being set to unity in the end.)

The zeroth order term is just the constant acceleration Rindler contribution. We shall now analyze the pole structure of the second term.
This term can be further simplified using expressions for $Q_{\pm}$. Specifically, the term in the round brackets in \eq{eq:poles-1} can be written, as
\begin{eqnarray}
Q_+ Q''_{-} - Q_- Q''_{+} = 2 \mathcal{A} \; Q_{-} - 2 \mathcal{B} \; Q_{+}
\end{eqnarray}
where
\begin{eqnarray}
\mathcal{A} &=& \l[ \xi \exp{+\xi} \r]_{\xi_1}^{\xi_2} - \frac{1}{2} \l[ \xi^2 \exp{+\xi} \r]_{\xi_1}^{\xi_2} \nn
\\
\mathcal{B} &=& \l[ \xi \exp{-\xi} \r]_{\xi_1}^{\xi_2} + \frac{1}{2} \l[ \xi^2 \exp{-\xi} \r]_{\xi_1}^{\xi_2}
\end{eqnarray}
with the obvious notation:
\begin{equation}
[\cdots]_{\xi_1}^{\xi_2} = [\cdots](\xi_2) - [\cdots](\xi_1).
\end{equation}
It can be shown that $\mathcal{A}$ and $\mathcal{B}$ are both finite at the poles. The second term in Eq.~(\ref{eq:poles-1}) therefore has cubic order poles determined by zeros of $Q_{\pm}$. Substituting the above expressions into Eq.~(\ref{eq:poles-1}), we obtain
\begin{eqnarray}
\frac{1}{I_+ I_-} \approx \frac{g_0^2}{Q_{+} Q_{-}} + \eta_0 \frac{g_0^2}{Q_{+} Q_{-}} \l[ \frac{\mathcal{A}}{Q_{+}} - \frac{\mathcal{B}}{Q_{-}} \r]
\label{eq:poles-2}
\end{eqnarray}
The first term is the standard Rindler contribution, and it is well known that this term gives a second order pole at $u_k = - i \beta_0 k$ with $k > 0$, and $\beta_0 = 2 \pi / g_0$. From here onwards, we shall denote derivatives with respect to $u$ by an overdot. We note that ${\dot Q}_{+} {\dot Q}_{-} / g_0^2 = 1$ at the poles, which is the standard result for Rindler and can be easily verified by explicit computation (usually, one uses the well known infinite image sum representation of $(\sinh{x})^{-2}$ to obtain the same result). The pole structure is now determined by
\begin{eqnarray}
Q_{+} Q_{-} Q_{\pm} = {\dot Q_{+}} {\dot Q_{-}} {\dot Q_{\pm}} \; \l( u - u_k \r)^3 + O\l(( u - u_k)^4 \r) \nn
\\
\end{eqnarray}
where the $u$ derivatives are to be evaluated at $u_k$. To compute the residues, we need to evaluate second derivatives with respect to $u$ of the functions $\mathcal{A} \exp{-i \omega u}$ and $\mathcal{B} \exp{-i \omega u}$, at $u=u_k$. This is straightforward and we relegate the details to Appendix \ref{app1}. The calculations are enormously simplified by noting that, as $t \rightarrow -t$, $\mathcal{B} \rightarrow \mathcal{A}$, so that we need to consider only terms \textit{odd} in $t$ in Eq.~(\ref{eq:poles-2}); the remaining terms (which would otherwise be tedious to evaluate), cancel.

The transition rate of the detector is given by Eq.~(\ref{eq:trans-rateF})
\begin{eqnarray}
\mathcal{\dot P} &=& \int_{-\infty}^{\infty} \DM u \exp{\l( -i \omega u \r)} \; G^{+}(u, t)
\nn \\
&=& - \frac{1}{4 \pi^2} \int_{-\infty}^{\infty} \DM u \; \frac{\exp{\l( -i \omega u \r)}}{I_{+} I_{-}}
\label{eq:trans-rate}
\end{eqnarray}
Substituting  the residues at the poles, calculated in Appendix \ref{app1}, we obtain
\begin{eqnarray}
\mathcal{\dot P} &=& \frac{1}{2 \pi} \sum_{k=1}^{\infty} \omega \exp{- \beta_0 \omega k}
\nn \\
&+& (\eta_0 t) \omega^2 \l[ 1 - \l( \frac{\pi}{\beta_0 \omega} \r)^2 \r] \sum_{k=1}^{\infty} k \exp{- \beta_0 \omega k}
\label{eq:res-app}
\end{eqnarray}
which is correct to $O(\eta_0)$. Hence, we see that the resultant spectrum will not be thermal at all frequencies even to order $O(\eta_0)$, due to the second term in the square brackets, which becomes significant at low frequencies. [A similar result was arrived at recently in \cite{townsend} in a different physical context.]

In the UV region (i.e., $\beta_0 \omega \gg 1$), we get
\begin{eqnarray}
\mathcal{\dot P} &=& \frac{1}{2 \pi} \sum_{k=1}^{\infty} \omega \exp{- \beta_0 \omega k}
\nn \\
&+& (\eta_0 t) \omega^2 \sum_{k=1}^{\infty} k \exp{- \beta_0 \omega k}
\end{eqnarray}
Noting that $\delta \beta = - 2 \pi \eta_0 t$, this can be written as
\begin{eqnarray}
\mathcal{\dot P} &=& \frac{1}{2 \pi} \l[ 1 + \delta \beta \frac{\partial}{\partial \beta} \r]_{\beta=\beta_0} \; \sum_{k=1}^{\infty} \omega \exp{- \beta \omega k}
\nn \\
&=& \frac{1}{2 \pi} \l[ 1 + \delta \beta \frac{\partial}{\partial \beta} \r]_{\beta=\beta_0} \l( \frac{\omega}{\exp{\beta \omega} -1} \r)
\end{eqnarray}
Therefore, to $O(\eta_0)$, we have
\begin{eqnarray}
\mathcal{\dot P} = \frac{1}{2 \pi} \frac{\omega}{\exp{\l[ \beta(t) \omega \r]} -1}
\approx\frac{1}{2 \pi} \omega\exp-[ \beta(t) \omega ]
\end{eqnarray}
where $\beta(t) = 2 \pi / g(t)$. This  result is intuitively understandable; at sufficiently high frequencies, we just recover the usual result with $g$ replaced by $g(\tau)$ when the acceleration varies with time.

However, note that it is valid only in the UV region; our result also shows that the spectrum will be modified for $\beta_0 \omega \ll 1$.
In fact, the second sum in Eq.~(\ref{eq:res-app}) is easily evaluated, and we obtain
\begin{eqnarray}
\mathcal{\dot P} = I_P[g_0] + \eta t \omega^2 \l[ 1 - \l( \frac{\pi}{s} \r)^2 \r]  \frac{\exp[s]}{\l[ \exp[s] - 1 \r]^2}
\end{eqnarray}
where, for convenience, we have defined $s=\beta_0 \omega$, and $I_P[g_0]$ represents Planck spectrum at temperature $g_0/(2 \pi)$. As stated above, for $s \gg 1$, the second term in square brackets can be neglected and the remaining terms combine to give $\mathcal{\dot P} \approx I_P[g(t)]$.

We want to analyze the $s \ll 1$ case a bit further, to highlight a  counter-intuitive fact. In this limit, we obtain,
\begin{eqnarray}
\mathcal{\dot P} &\approx& I_P[g_0] - \eta t \omega^2 \l( \frac{\pi}{s} \r)^2 \l[ \frac{1}{s^2} - \frac{1}{12} + O(s^2) \r]
\nn \\
&\approx& I_P[g_0] - \eta t \omega^2 \l( \frac{\pi}{s^2} \r)^2
\end{eqnarray}
In the same limit, $I_P[g_0] \approx \omega / (2 \pi s) = C$ (say), so that we can rewrite the above expression as
\begin{eqnarray}
\mathcal{\dot P} &\approx& C \l[ 1 - (2 \pi^3) \frac{\eta}{s^2} \frac{\omega t}{s} \r]
\nn \\
&=& C \l[ 1 - (2 \pi^3) b a^2 {\omega t} \r]
\label{eq:dist-tmp}
\end{eqnarray}
where $b$ and $a$ are the dimensionless quantities,
\begin{equation}
a=1/s=g_0/2\pi \omega;\;b=\dot g_0/(g_0 \omega)
\end{equation}

Evidently, there are two possibilities, depending on whether $b$ is greater than or less than one. When
\begin{eqnarray}
&s& \ll \eta \ll 1, \mathrm{~or}
\nn \\
&1& \ll b \ll a
\end{eqnarray}
we see that the frequency is probing the change in acceleration because $\omega^{-1}\gg (\dot g/g)^{-1}$. So we certainly expect the spectrum to be distorted and this is what happens and
the result is understandable. However, when
\begin{eqnarray}
&\eta& \ll s \ll 1, \mathrm{~or}
\nn \\
&b& \ll 1 \ll a
\end{eqnarray}
we see that
$\dot g / g \ll \omega$ and so these frequencies are not probing the change in acceleration at all. Therfore, one would have expected to recover the results of constant acceleration in which case there are no distortions from the thermal spectrum at any frequency. But we see that, in this case, we can still have $ba^2 \ll 1$ and produce a distortion of thermal spectrum at low frequencies.

Before proceeding further, we must highlight an important assumption that has gone into the derivation. For
calculating the residues, we have first expanded the integrand and then evaluated the residues. The true expression
is, of course, to be obtained by first doing the contour integral and then expanding in $\eta$. We have
assumed that the two steps, Taylor expansion and integration, commute. Our result will be invalidated for
functions $g(\tau)$ which fail to satisfy this criterion. Moreover, we have also assumed that the $u$ integration
goes all the way from $-\infty$ to $+\infty$, while doing a Taylor series in $t$. It is important to understand
better whether such an approximation is valid, and, if not, what difference will it make to the result. In
particular, the low frequency modification we obtain may be an artifact of such a truncation, and this caveat must
always be kept in mind.

\section{Detector response for a Specific trajectory } \label{sec:specific-acc}
We shall now study the response function corresponding to a particular detector trajectory determined by
\begin{eqnarray}
g(\tau) = \frac{g_0}{1 + \epsilon g_0 |\tau|}
\end{eqnarray}
where $\epsilon$ is a small, dimensionless parameter. The response can now be evaluated in a straightforward manner. We expect a splitting of the quadratic poles at $O(\epsilon)$ from their constant acceleration values, so that we essentially have a couple of first order poles separated infinitesimally. For the above $g(\tau)$, we have
\begin{equation}
\chi(\tau)= (1/\epsilon) \ln{(1+\epsilon g_0 |\tau|)} \rm{sgn}{(\tau)}.                                                                         \end{equation}
Because of the dependence on $|\tau|$, we need to consider the cases (i) $0<\tau_1<\tau_2$, (ii) $\tau_1<\tau_2<0$ and (iii) $\tau_1<0, \tau_2>0$ separately while evaluating $I_{\pm}$ (we take $\tau_2>\tau_1$ without loss of generality). While the first two cases admit analytic expressions for the poles, the same is not true of (iii). We shall evaluate the response function for the first two cases, and comment on the possible effect of (iii) later on.

The integrals involved in $I_{\pm}$ are trivial; for clarity, we refer to values of $I_{\pm}$ for case (ii) as $I^N_{\pm}$, and those for case (i) simply as $I_{\pm}$. Then, it is easy to see that $I^N_{\pm}(\tau_1, \tau_2) = I_{\mp}(-\tau_2,-\tau_1)$. So, we can obtain $I^N_{\pm}$ from $I_{\pm}$ simply by changing $t$ to $-t$ (see definitions of $u$ and $t$ above), or, what is the same thing, by replacing $t$ with $|t|$ in the expression for case (i). With this understanding, we simply write $t$ rather than $|t|$ in the expressions below. We also define $\xi_{1(2)} = 1 + \epsilon g_0 \tau_{1(2)}$ and $\eta=\xi_2/\xi_1$. Then,
\begin{equation}
I_{\pm} = (\xi_1/g_0) (\epsilon \pm 1)^{-1} \l[ \eta \; \xi_2^{\pm (1/\epsilon)} - \xi_1^{\pm (1/\epsilon)} \r].                                                                                                                \end{equation}
The poles (i.e., the zeros of $I_{\pm}$) are determined by: $(1 \pm \epsilon^{-1}) \ln{\eta} = 2 \pi i k$. Now rewrite $\xi_1$ and $\xi_2$ in terms of $u$ and $t$, to obtain
\begin{eqnarray}
u_k^{\pm} = \frac{2 i}{g_0} \l( \epsilon^{-1} + g_0 t \r) \tan{\l( \frac{\chi_k^{\pm}}{2} \r)}
\label{eq:pole-sp}
\end{eqnarray}
where $\chi_k^{\pm} = 2 \pi k / ( 1 \pm \epsilon^{-1} )$. As a check, note that this reduces to the standard constant acceleration values, $\pm 2 i \pi k / g_0$ for $\epsilon=0$.

Rest of the calculation involves standard residue calculus, and is quite lengthy. Since the poles are now split at $O(\epsilon)$, we need to evaluate the quantities $R_+=I'_{+} (u_k^+) I_-(u_k^+)$ and $R_-=I'_{-} (u_k^-) I_+(u_k^-)$ for calculating the residues. The quantity $R_{\pm} \times g_0 / (1+\epsilon  g_0 t)$ is given by
\begin{eqnarray}
\frac{1+\exp\left[\pm i \frac{\chi_k^{\pm}}{\epsilon }\right]-\exp\left[i \chi_k^{\pm} \left(1 \mp \frac{1}{\epsilon }\right)\right]-\exp[i \chi_k^{\pm}]}{\l(-\epsilon \pm 1\r) \l(1+\exp[i \chi_k^{\pm}\r)]}
\nn \\
\label{eq:res-sp}
\end{eqnarray}
It is now a straightforward exercise to use Eqs.~(\ref{eq:pole-sp}) and (\ref{eq:res-sp}) and evaluate the response function. This turns out to be
\begin{eqnarray}
\mathcal{\dot P} = \frac{1}{2 \pi} \sum_{k=1}^{\infty} \l( \omega - \epsilon 2 \pi \omega^2 t k + O(\epsilon^2) \r) \exp{\l(-\frac{2 \pi}{g_0} \omega k\r)}
\end{eqnarray}
No further calculations are required, since it easy to see that, with $\delta g = - \epsilon g_0^2 t + O(\epsilon^2)$, the two terms above combine to give a Planck spectrum with temperature $g(t)/2 \pi$.

Let us now turn to the contribution of poles which we have not accounted for. In the above calculation, we left out the contribution to the integral of the $u$-range where $\tau_1, \tau_2$ have opposite signs. Unfortunately, this case does not admit analytic expressions for the poles. But, from the result in Sec. \ref{sec:time-dep-acc}, we expect that this contribution will be   irrelevant at high frequencies. Apart from this, it is not possible to make any comments about this contribution. As already mentioned in the Introduction, this is typical of most of the calculations that attempt a rigorous evaluation [in particular, \cite{louko-satz} discusses characteristics of detector response in curved spacetime] although the explicit result we have obtained is very close to what one would have expected for a slowly changing acceleration.
\section{Conclusions} \label{sec:comments}

The result brings out another interesting fact associated with the combined effect of presence of the horizon and varying acceleration and we shall discuss this briefly.

We see that, for modes with $\omega^{-1} \ll g^{-1}$, the result essentially involves replacing the acceleration by its instantaneous value so that $T(\tau)=g(\tau)/2\pi$ but the thermal spectrum gets distorted for $\omega^{-1} \gg g^{-1}$. At first sight, one would have thought that this is to be expected. We know that for an accelerated trajectory, $g^{-1}$ gives the approximate location of the local horizon. (The exact location of the horizon will change with $\tau$, see Appendix \ref{app2}).  On the other hand, a mode with frequency $\omega$ will probe a length scale $\sim \omega^{-1}$. Such a mode will be within the region `outside the horizon'  if $\omega^{-1} \ll g^{-1}$, or $\beta \omega \gg 1$ but will probe the horizon scale and beyond if $\omega^{-1} \gg g^{-1}$. So one may think that it is natural for the spectrum to be distorted in the latter case.

There is, however, a subtlety here. Our problem actually has \textit{three} length scales not just two: $\omega^{-1}, g^{-1}$, and quite \textit{crucially}, $(\dot g/g)^{-1}$. It would have been no surprise, if  the spectral distortion arose for $\omega^{-1}\gg(\dot g/g)^{-1}$; these are the frequencies which see the change in the acceleration and there will be some distortion. But this is not what we found!
Instead we find that in a typical situation with $g^{-1}\ll (\dot g/g)^{-1}$, with very slowly varying acceleration, the spectral distortions occur already when $g^{-1}\ll\omega^{-1}\ll
(\dot g/g)^{-1}$. So whether $\dot g\neq0$ or whether $\dot g=0$ makes a difference to the spectrum even when the relevant frequency is not probing the time variation of the acceleration. Obviously, this effect does not exist in the case of $\dot g=0$; so we need to conclude that one cannot take the limit continuously for all frequencies. We believe this arises due to the changing distance to the horizon but only further investigations will nail down the precise reason.

It is known that viewed from the inertial frame the final state of the field is an one-particle state so that the `detection' is actually accompanied by an emission. It has been suggested [see \cite{tpclick}] that it is better to think of the detector as radiating a Minkowski particle, rather than ``detecting" anything. From this point of view, it would be interesting to see whether the response function we have obtained has some simple interpretation, particularly the $\dot g$ term.

Finally, as promised in the Introduction, we briefly discuss the issue of comparing our results with other results in
literature. In doing so, one must realize that our expressions for the rate as well as the regularization are
different from the ones used in the literature. Our choices are based on the simple fact that it is closest to what one
does in the standard, constant acceleration case. A different choice of variables for defining the rate (and even a
different regularization scheme) can alter the results since the pole structure will change. In this context, we would
particularly like to mention the analysis presented in \cite{louko-satz, milgrom}
Our choice of regularization (with an $i \epsilon$ prescription on $u$) is actually similar to the one employed in
\cite{milgrom} (see their Eq.~(21)). However, the difference lies in the definition of the response rate itself. In
particular, the relevant function which gives transition rate at time $t$, for e.g., in Refs. \cite{louko-satz} and
\cite{milgrom}, is $G(t,t-u)$. It is not difficult to see that this would have a completely different functional
dependence on $t$ and $u$ as compared to
our case, since the definition of $t$ is manifestly different. So, effectively one is integrating completely
different functions of $u$ in the two cases; the results can not, therefore, be directly compared as such.
Additional physical criteria are needed to choose one definition over another (for e.g., in \cite{louko-satz} and
\cite{milgrom}, the motivation is causality. In our case, the coordinates $t$ and $u$ corresponding to the two points
on the trajectory actually correspond to the so called ``radar coordinates" which are natural set of local coordinates
assigned to nearby points connected to $(\tau_1, \tau_2)$ by light beams. By their very construction, these
coordinates are non-local, and the transition rate must be interpreted keeping this in mind. However, further work is
needed to make a precise connection.)

\section*{ACKNOWLEDGEMENTS}
We would like to thank Aseem Paranjape for carefully going through the manuscript, and giving several useful comments.
We are also grateful to Prof. Jorma Louko for some enlightening comments on the result. DK is supported by a
Fellowship from the Council of Scientific \& Industrial Research (CSIR), India.
\appendix
\section{Evaluation of residues of Eq.~(\ref{eq:trans-rate})} \label{app1}
In this appendix, we outline the evaluation of residues at the cubic order poles of Eq.~(\ref{eq:trans-rate}). We would only stress certain steps which are crucial to minimize the algebra in the otherwise elementary calculation.

To begin with, define two new variables, $a=g_0 u / 2$ and $b=g_0 t$, so that, $\xi_1 = b-a$ and $\xi_2=b+a$. We then have
\begin{eqnarray}
\mathcal{A} &=& 2 \; {\rm e}^{+b} \cosh{a} \l[ a(1-b) + \tanh{a} \l( +b - \frac{1}{2} b^2 - \frac{1}{2} a^2 \r) \r]
\nn \\
\mathcal{B} &=& 2 \; {\rm e}^{-b} \cosh{a} \l[ a(1+b) + \tanh{a} \l( -b - \frac{1}{2} b^2 - \frac{1}{2} a^2 \r) \r]
\nn
\\
\label{eq:tmp-A1}
\end{eqnarray}
It is evident that $B(a, b) = A(a, -b)$, as mentioned in the text. We essentially require [see Eq.~(\ref{eq:poles-2})] $\mathcal{A} / {\dot Q}_+$ and $\mathcal{B} / {\dot Q}_-$. From their definition (\ref{eq:Q-def}), we have
\begin{eqnarray}
{\dot Q}_{\pm} = g_0 \; {\rm e}^{\pm b} \cosh{a}
\label{eq:tmp-A2}
\end{eqnarray}
Note that ${\dot Q}_{+} {\dot Q}_{-} / g_0^2 = 1$ at the poles, which, as was emphasized in the text, is the standard result for Rindler. So, we have $\l[ \mathcal{A} / {\dot Q}_{+} \r](a, b) = \l[ \mathcal{B} / {\dot Q}_{-} \r](a, -b)$. Therefore, the contribution of the $O(\eta_0)$ term in Eq.~(\ref{eq:poles-2}) to $\mathcal{\dot P}$ becomes
\begin{eqnarray}
\l(-\frac{1}{4 \pi^2} \r) \times \l( \frac{-2 \pi i}{\;\; 2 !} \r) \frac{\eta_0}{\dot Q_{+}(u_k)} \l[ \frac{\DM^2}{\DM u^2} \l( \mathcal{A} \exp{-i \omega u} \r) \r]_{u=u_k}
\nn
\\
\label{eq:tmp-A3}
\end{eqnarray}
plus a similar term for $\mathcal{B}$, and a sum over all relevant $k$'s (such that $u_k$'s lie on the negative imaginary axis, with the contour closed in the lower-half complex $u$-plane). All that remains is to pick out the terms in $\mathcal{A}$ which are odd in $b$, calculate the second derivatives which are required, evaluate at $u_k$, and then multiply by $2$ for the contribution of the $\mathcal{B}$ part. These are all straightforward, though lengthy, steps. The object of interest is [see Eqs.~\ref{eq:tmp-A1}, \ref{eq:tmp-A2} and \ref{eq:tmp-A3}],
\begin{equation}
f = e^{-b} \mathcal{A} = 2 (\cosh{a}) \l[ b \l( \tanh{a} - a \r) \r] + (\text{terms even in} \,b),
\end{equation}
and, at the poles, we obtain,
\begin{equation}
\dot f = 0, \ddot f = - 2 a_k {\dot a}_k^2 b \cosh{a_k} \ \text{and}\ f=-2 a_k b \cosh{a_k}. \end{equation}
Putting everything together, we finally obtain Eq.~(\ref{eq:res-app}). [Note that the second term in the square brackets in Eq.~(\ref{eq:res-app}), which becomes significant in IR, arises from the $\ddot f$ term above.]
\section{Local horizon for a trajectory with time dependent acceleration} \label{app2}
In the local coordinates based on a trajectory with time-dependent acceleration [see Eq.~(\ref{eq:local-coords})], the metric becomes (see for e.g, Ref. \cite{tppr})
\begin{eqnarray}
\DM s^2 = - \l[ 1 + g(\tau) x \r]^2 \DM \tau^2 + \DM x^2 + \DM Y^2 + \DM Z^2
\end{eqnarray}
The equation for a null surface can be written as $\Phi(\tau, x) = x - f(\tau) = 0$. The function $f(\tau)$ is determined by the condition $\partial_a \Phi \partial^a \Phi = 0$. Doing this leads to a differential equation for $f(\tau)$; it's solution yields the following expression for the horizon location
\begin{equation}
x_{\rm H}(\tau) = p \exp{- \xi(\tau)} \int^{\tau} \DM y \; \exp{\xi(y)}
\end{equation}
where
\begin{equation}
\xi(\tau) = \mp \int^{\tau} g(x) \DM x
\end{equation}
For $g(\tau) \approx g_0 + \dot g_0 \tau$, we have
\begin{equation}
x_{\rm H}(\tau) = - g_0^{-1} \l[ 1 - \eta ( \pm 1 + g_0 \tau ) + O(\eta^2) \r]
\end{equation}
We can also invert this to write, to the same order of accuracy,
\begin{equation}
g(\tau) \simeq - x_{\rm H}^{-1} \l[ 1 \mp \; \eta \r]
\end{equation}
The temperature $T(\tau) = g(\tau) / 2 \pi $, associated with the detector response at $O(\eta)$ and in the UV limit, can be cast in an interesting form by further noting that, $v = \dot x_{\rm H} = \eta + O(\eta^2)$:
\begin{eqnarray}
T(\tau) \simeq \frac{g(\tau)}{2 \pi} \simeq - \frac{1}{2 \pi x_{\rm H}}  \frac{1}{1 \pm v}
\end{eqnarray}



\begin{thebibliography}{99}
\bibitem{comment}
Incidentally, this can also be seen directly  in the Euclidean sector, in which the hyperbolic trajectory of the uniformly accelerated detector maps to a circle of constant radius $g_0^{-1}$. Now, $G^+$ for any two points on the circle depends on the chordal distance between the points, and it follows from trivial geometry that this chordal distance can be completely expressed in terms of $\sin(\Delta \theta)$ where $\Delta \theta$ is the angular separation between the points. Analytically continuing back, we see that $G^+$ depends only on $u=i \Delta \theta$.
\end{thebibliography}
\end{document}